\documentclass[prb,twocolumn,groupedaddress,shownopacs,amssymb]{revtex4-2}
\usepackage{graphicx,psfrag,amsmath,amssymb}
\usepackage[usenames, dvipsnames]{color}
\usepackage{braket}
\usepackage{amsmath}
\usepackage{epsfig}
\usepackage{subfigure}
\usepackage{mathtools}
\usepackage{setspace}
\usepackage{bm}
\usepackage{ulem}
\usepackage{times}
\usepackage{enumitem}
\usepackage{xcolor}
\usepackage{multirow}
\usepackage{soul}
\usepackage{dcolumn}
\usepackage{balance}
\usepackage{epstopdf}
\usepackage{braket}
\usepackage{color}
\usepackage{diagbox}
\usepackage{lineno}
\usepackage[utf8]{inputenc}

\usepackage{hyperref}
\hypersetup{
    colorlinks=true,
    linkcolor=Blue,
    citecolor=Blue,
    filecolor=Blue,
    urlcolor=Blue,
    }
\def\Tr{\mathrm{Tr}}

\begin{document}
%\pagewiselinenumbers
\title{Boundary-driven many-body phase transitions in a non-Hermitian
       disordered fermionic chain}
\author{Kuldeep Suthar}
\affiliation{Department of Physics,  
             Central University of Rajasthan, Ajmer - 305817, India}
\date{\today}

%%%%%%%%%%%%%%%%%%%%%%%%%%%%%%%%%%%%%%%%%%%%%%%%%%%%%%%%%%%%%%%%%%%%%%%%%%%%%%
%%%%%%                             Abstract                               %%%%
%%%%%%%%%%%%%%%%%%%%%%%%%%%%%%%%%%%%%%%%%%%%%%%%%%%%%%%%%%%%%%%%%%%%%%%%%%%%%%
\begin{abstract}
  The non-Hermitian systems exhibit extreme sensitivity to the boundary 
conditions. The change in the eigenspectrum with tunning boundary parameter is 
intimately connected to the non-Hermitian skin effect. The single-particle 
systems are affected by the boundary perturbations; however the interplay of a 
random disorder potential and non-reciprocal hopping under boundary 
perturbations of an interacting many-body system is not yet clear. In this 
work, we examine the boundary sensitivity of a non-Hermitian interacting 
fermionic system in the presence of a random disorder potential. A non-zero 
boundary parameter results in real-complex spectral transitions with 
non-reciprocal (or unidirectional) hopping at weak disorder. While the 
many-body localization at strong disorder washes away real-complex transitions 
leading to dynamical stability and real eigenvalue spectrum. We show that the 
boundary-driven real-complex spectral transitions of the non-Hermitian chain 
are accompanied by the corresponding changes in the level statistics and 
nearest level-spacing distributions. The intriguing features of non-reciprocity
and boundary sensitivity are further revealed using the averaged inverse 
participation ratios. Finally, we find distinct behaviour in the quench 
dynamics of local particle density, population imbalance, and entanglement 
entropy of charge-density-wave ordered state that corroborate the real-complex 
and localization transitions. Our results provide a route to understanding 
disordered many-body systems under a generalized boundary. 
\end{abstract}

\maketitle 
%%%%%%%%%%%%%%%%%%%%%%%%%%%%%%%%%%%%%%%%%%%%%%%%%%%%%%%%%%%%%%%%%%%%%%%%%%%%%%%
%%%%                           I. Introduction                             %%%% 
%%%%%%%%%%%%%%%%%%%%%%%%%%%%%%%%%%%%%%%%%%%%%%%%%%%%%%%%%%%%%%%%%%%%%%%%%%%%%%%
\section{Introduction}
  The investigations of the physical properties and topological phases of 
non-Hermitian systems have gained significant attention in recent 
years~\cite{ashida_20,bergholtz_21,weidemann_22,kawabata_23,yue_23,yang_24}. 
These systems are non-conservative in which the inherent gain or loss emerges 
due to coupling with external environments~\cite{gong_18,li_24}. The rapid 
technical advances in the ultracold atomic gases~\cite{jiaming_19,lapp_19,
zejian_22,liang_22,medina_22,dong_24}, electrical 
circuits~\cite{imhof_18,liu_23}, and 
photonic~\cite{liang_14,khanikaev_17,yang_18,lu_23} and acoustic 
systems~\cite{yang_15,xue_22,lu_23} provide ideal platform to realize the novel 
phenomena in experiments. Moreover, the absence of thermalization in isolated 
disordered interacting Hermitian system can results in a phenomenon dubbed 
as~\textit{many-body localization} (MBL)~\cite{alet_18,abanin_19,sierant_24}. 
The successful attempts to engineer the controllable environment pave a way to 
explore the effects of the inevitable environment on the localization 
properties of interacting disordered systems. These advancements provide an 
opportunity to discover new classes of transitions beyond the realm of 
Hermitian critical phenomena. 

In the absence of hermiticity, the eigenspectrum of the system could be complex,
which leads to many intriguing phenomena in non-Hermitian systems. One of the 
remarkable phenomenon is the \textit{non-Hermitian skin effect} 
(NHSE)~\cite{zhang_22} which refers to an anomalous localization of extensive 
eigenmodes at the boundaries. Recently, the non-Hermitian Hamiltonians with 
dissipations and non-reciprocal hopping have been realized in 
experiments~\cite{jiaming_19,gou_20,zejian_22,liang_22}, where the 
characteristic signature of NHSE and topological phenomena, including 
geometry-dependent~\cite{wang_23a} and high-order NHSE~\cite{zhang_21} have 
been observed in experiments. The existence of NHSE signals the breakdown of 
bulk-boundary correspondence~\cite{kunst_18,yao_18,halder_23}. The energy 
spectrum becomes highly sensitive to the boundary conditions as the open 
boundary condition (OBC) and periodic boundary condition (PBC) show distinct 
features. Various versions of NHSE have recently been 
proposed~\cite{kawabata_20,borgnia_20,gu_22,lin_23,yoshida_24,ma_24}. Previous 
studies on the boundary sensitivity primarily dealt with the level of 
single-particle systems and revealed the interplay of system size and boundary 
perturbations on the existence of NHSE. The NHSE can exist even beyond OBC when
one of the hopping terms is absent~\cite{guo_21}. In recent times, the role of 
many-body correlations on the properties of non-Hermitian systems, including 
the nontrivial topology~\cite{zhang_22a}, relaxation dynamics~\cite{zhang_20}, 
and entanglement transition~\cite{lu_24,liu_24}, have been examined. However, 
the investigations of NHSE for interacting disordered systems are limited. The 
non-Hermitian many-body localization in 
disordered~\cite{hamazaki_19,suthar_22,mak_24}, 
quasiperiodic~\cite{zhai_20,wang_23,cheng_24}, and electric-field-driven stark 
potential~\cite{li_23,liu_23a} have been proposed. The two localization 
phenomena, NHSE and MBL, of the non-Hermitian interacting chain can be 
distinguished by the eigenstate properties and non-equilibrium quench 
dynamics~\cite{wang_23}. Despite several theoretical investigations on 
non-Hermitian systems, the role of the boundary perturbations under the 
presence of random disorder potential for interacting non-Hermitian fermionic 
systems is yet to be examined. 

In the present work, we explore the boundary sensitivity in the emergence of 
real-complex spectral and localization transitions of a many-body non-Hermitian
disordered chain. Here, the spectral transition refers to the real-complex 
transition of the eigenenergies by a change in the boundary parameter or 
disorder strength in a single fermionic chain. The coupling between the two end
sites of the chain is controlled by a boundary parameter. We first examine the 
eigenvalue spectra due to an interplay of hopping with boundary parameters in a
one-dimensional interacting chain subjected to a random disorder. We find that 
the real-complex transitions are accompanied by localization transitions. The 
localization properties of the system are examined using the inverse 
participation ratio and nearest-neighbour level statistics. We demonstrate that
the non-reciprocal hopping results in the coexistence of real-complex and 
localization transitions, and non-Hermitian many-body localization at strong 
disorder dynamically stabilizes the system. Further, we perform the quench 
dynamics of a density-wave ordered initial state and analyze the 
disorder-averaged local particle density, occupancy imbalance, and the 
entanglement entropy. The localization of particles at the chain boundaries is 
characterized by the time-dynamics of particle density and imbalance, and 
entanglement entropy signals the signature of many-body localization in the 
non-reciprocal (and reciprocal) lattice with boundary tuning parameter. 

The rest of this paper is organized as follows: Section~\ref{model_ham} 
introduces the Hatano-Nelson model with variable boundary terms. In 
Section~\ref{results}, we discuss the real-complex transition, level 
statistics, inverse participation ratio, and quench dynamics of dynamical 
observables. Finally, the findings are summarized in Section~\ref{conc}.

%%%%%%%%%%%%%%%%%%%%%%%%%%%%%%%%%%%%%%%%%%%%%%%%%%%%%%%%%%%%%%%%%%%%%%%%%%%%%%%
%%%%%                      II. Model Hamiltonian                          %%%%%
%%%%%%%%%%%%%%%%%%%%%%%%%%%%%%%%%%%%%%%%%%%%%%%%%%%%%%%%%%%%%%%%%%%%%%%%%%%%%%%
\section{Model Hamiltonian}
\label{model_ham}
We consider a one-dimensional Hatano-Nelson 
chain~\cite{hatano_96,hatano_97,hatano_98} of interacting fermionic atoms 
loaded in a random disorder potential with generalized boundary conditions. 
The model Hamiltonian of the system is
\begin{align}
  \hat{H} =&-\sum_{j=1}^{L-1} \left(J_L \hat{c}^{\dagger}_{j} 
	     \hat{c}_{j+1}
            + J_R \hat{c}^{\dagger}_{j+1} \hat{c}_{j}
	     \right) 
	    + \delta_R \hat{c}^{\dagger}_{1} 
	     \hat{c}_{L} + \delta_L \hat{c}^{\dagger}_{L} 
             \hat{c}_{1} 
             \nonumber  \\ 
	   &+ V \sum_{j} \hat{n}_{j} \hat{n}_{j+1} 
	    + \sum_{j} \epsilon_{j} \hat{n}_{j}.
\label{model}
\end{align}
Here, $j$ represents the spatial lattice index and $L$ is the number of lattice 
sites, $J_{L(R)}$ denotes the left (right)-hopping amplitude for neighbouring 
lattice sites, $\delta_L$ and $\delta_R$ determine generalized boundary 
conditions, $V>0$ is a repulsive coulomb interaction strength between two 
spinless fermions at adjacent sites, and $\epsilon_{j}$ is a random disorder 
potential chosen between $[-W,W]$ with $W$ being the disorder strength. Here, 
we have considered $\delta_L=\delta_R=\delta$ in the present work. In the 
absence of a disorder potential, the single-particle limit of the model 
displays a system-size dependent boundary effects and real-complex transitions 
of eigenspectra~\cite{guo_21}. For $V=W=0$, the eigenvalue equation 
corresponding to the above Hamiltonian is 
$\hat{H}_{\rm sp}\ket{\Psi} = E_{\rm sp}\ket{\Psi}$, where the eigenstate 
$\ket{\Psi} = \sum_{j}\psi_{j}\ket{j}$ with 
$\ket{j} = \hat{c}^{\dagger}_{j}\ket{0}$ and $\hat{H}_{\rm sp}$ is a 
single-particle Hamiltonian corresponding to Eq.~(\ref{model}). The solutions 
of the eigenvalue equation are related by the following relation~\cite{guo_21}
\begin{equation}
  \sin[(L+1)\theta] - \xi_{1}\sin[(L-1)\theta] - \xi_{2}\sin[\theta] = 0, 
  \label{condition}
\end{equation}
where $\xi_{2} = \left({\delta_{L}}/{J_{L}}\right)
\left(J_{R}/J_{L}\right)^{-L/2} + \left(\delta_{R}/J_{R}\right)
\left(J_{R}/J_{L}\right)^{L/2}$ and $\xi_{1} 
= \left({\delta_{L}\delta_{R}}/{J_{L}J_{R}}\right)$. 
The solutions $\theta$ of the above equation depends on hopping amplitudes and
boundary terms through $\xi_{1}$ and $\xi_{2}$, and can in general possess both
real and complex values. Once the boundary terms are finite, the solutions of 
the equations become highly sensitive to the system size. The finite NHSE can 
exist when the ratio of boundary term to one of the hopping amplitudes 
$(\delta_{L}/J_{L})$ (at fixed $(\delta_{R}/J_{R})$) is smaller than 
$\approx 0.07$ with $J_{L}\neq J_{R}$. In case of open chain with 
$\delta_{L}=\delta_{R}=0$, which leads to $\xi_1 = \xi_2 = 0$, and the above 
Eq.~(\ref{condition}) gives $L$ real solutions $\theta = j\pi/(L+1)$. 
Consequently, the eigenvalues of the system are real, and the eigenstates take 
the form $\Psi = \left( r\sin[\theta], r^{2}\sin[2\theta], 
...... ,r^{L}\sin[L\theta]\right)^{T}$, where $r=\sqrt{J_{R}/J_{L}}$. Moreover, 
at higher ratios of $\delta$ to hopping amplitudes, the skin effect vanishes, 
and the eigenspectrum of the system exhibits a real-complex transition. It is 
important to note that the above model Hamiltonian [Eq.~(\ref{model})] 
possesses real-complex transitions as it respects the time-reversal 
symmetry~\cite{hamazaki_19,suthar_22}. We discuss in the next section the 
numerical results of real-complex and non-Hermitian localization transitions in
the presence of a disorder potential.

%%%%%%%%%%%%%%%%%%%%%%%%%%%%%%%%%%%%%%%%%%%%%%%%%%%%%%%%%%%%%%%%%%%%%%%%%%%%%%%
%%%%%                 III. Results and Discussions                        %%%%%
%%%%%%%%%%%%%%%%%%%%%%%%%%%%%%%%%%%%%%%%%%%%%%%%%%%%%%%%%%%%%%%%%%%%%%%%%%%%%%%
\section{Results and Discussions}
\label{results}
 In this section, we first discuss our numerical results of static properties 
such as real-complex transitions of eigenenergies, and diagnostics of 
localization and extended character followed by dynamical properties. To reveal
the interplay of non-hermiticity and boundary sensitivity, we consider the 
generalized boundary conditions on non-reciprocal couplings and numerically 
study the effects of random disorder potential. In the present study, the 
on-site interaction strength is fixed at $V=1$ and half-filling case is 
considered. The static and dynamical properties are primarily examined for the 
system size $L=16$.

%%%%%%%%%%%%%%%%%%%%%%%%%%%%%%%%%%%%%%%%%%%%%%%%%%%%%%%%%%%%%%%%%%%%%%%%%%%%%%%
%%%%%                       A. Spectral transition                         %%%%
%%%%%%%%%%%%%%%%%%%%%%%%%%%%%%%%%%%%%%%%%%%%%%%%%%%%%%%%%%%%%%%%%%%%%%%%%%%%%%%
\subsection{Real-complex transition of eigenenergies}
\label{spec_tran}
 We first study the boundary effects on the complex energy spectra as a 
function of the disorder potential strength. Various regimes can be 
distinguished based on the fraction of the complex energies averaged over 
several disorder realizations. The fraction of the complex energies is defined 
as $f_{\rm im}=\overline{D_{\rm im}/D}$~\cite{hamazaki_19,zhai_20,suthar_22}, 
where $D_{\rm im}$ is the number of eigenenergies whose imaginary part 
$|{\rm Im}\{E\}|\geqslant C$ with a cut-off $C=10^{-13}$, which is identified 
as the error in our numerical diagonalization. Here, $E$ is the complex 
eigenenergies and $D$ is the total number of eigenenergies. The overline 
denotes the average over several disorder samples. 
\begin{figure}[ht]
  \includegraphics[width=\linewidth]{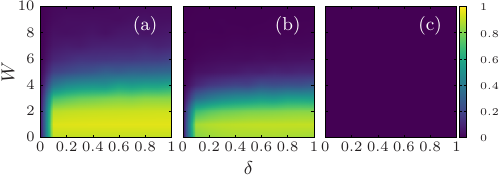}
  \caption{The disorder-averaged complex energy fraction as a function of
           boundary parameter and disorder strength. Here $J_L$ is varied to 
	   reveal the non-reciprocal coupling effects with $J_R=1$. 
	   (a) $J_L = 0.0$, (b) $J_L = 0.5$, and (c) $J_L = 1.0$. The system 
	   size is $L=12$. The fraction shown is averaged over $300$ disorder 
	   realizations.}
\label{imag-frac}
\end{figure}

The spectrum of the non-Hermitian many-body lattice model [Eq.~(\ref{model})] 
shows strong sensitivity to the boundary parameter. For a single-particle clean 
system, the eigen-spectrum describes a loop in the complex energy plane as 
$\delta$ increases. The region of complex loop structured energy spectra 
enhances as $L$ increases while the real spectra region becomes 
narrow~\cite{guo_21}, which is consistent with the analytical solutions of 
Eq.~(\ref{condition}). For many-body clean systems, the spectrum follows that 
of larger single-particle systems. Moreover, it is important to note that the 
single-particle eigenstates get delocalized (at the band center) due to pinning
of the flux lines in the presence of non-reciprocal hoppings. The random 
disorder potential leads to a statistically symmetric spectrum in the complex 
plane and localizes the eigenstates. Thus, the uncorrelated disorder results in
a real spectrum even for single-particle 
systems~\cite{hatano_96,hatano_97,hatano_98,hatano_21}. Fig.~\ref{imag-frac} 
shows $f_{\rm im}$ as a function of the boundary parameter $\delta$ and 
disorder strength $W$. The value of $f_{\rm im}$ is averaged over several 
disorder realizations in the presence of random disorder. We set one of the 
hoppings $J_{R}=1$. First, we examine the energy spectrum for unidirectional 
(non-reciprocal) hopping case in which $J_{L}=0$ [Fig.~\ref{imag-frac}(a)], 
where $f_{\rm im}=0$ and $f_{\rm im}=1$ represents completely real and complex 
spectrum. At $\delta=0$, the model corresponds to an open chain and can be 
mapped to a Hermitian system through imaginary gauge 
transformation~\cite{hatano_21,suthar_22}. This results in a real spectrum that
remains valid even with finite disorder. Once a finite value of $\delta$, a 
link connecting both ends of the boundary, is introduced the generalized 
boundary leads to complex spectrum at weak disorder. This is evident from a 
sharp change in $f_{\rm im}$ at the boundary for $\delta=0$ and $\delta=0.1$. 
As $W$ increases, the $f_{\rm im}$ lowers and finally becomes zero (real 
spectrum) at a critical disorder strength. The distributions of the eigenenergy
spectrum suggest that the critical $W$ increases as $\delta$ approaches one. 
The strong disorder suppresses the imaginary parts of the complex energies and 
dynamically stabilizes the system~\cite{hamazaki_19,suthar_22}. The genesis of 
the realness of the spectrum lies at the single-particle level.

We further find that non-reciprocal hopping with $J_{L}=0.5$ exhibits a similar
eigenenergy spectrum; however, the complex energy regime is reduced. Hence, the 
critical $W$ of complex-real transition depends on the difference of hopping in
left and right directions. Once both hoppings are equal which turn into the 
Hermitian limit, the $\delta-W$ plane possess real spectrum. This demonstrates 
the boundary sensitivity is applicable for systems with non-reciprocal hopping.

%%%%%%%%%%%%%%%%%%%%%%%%%%%%%%%%%%%%%%%%%%%%%%%%%%%%%%%%%%%%%%%%%%%%%%%%%%%%%%%
%%%%%                      B. Inverse participation ratio                  %%%%
%%%%%%%%%%%%%%%%%%%%%%%%%%%%%%%%%%%%%%%%%%%%%%%%%%%%%%%%%%%%%%%%%%%%%%%%%%%%%%%
\subsection{Inverse participation ratio}
\label{ipr_sec}
We now examine the localization properties of the non-Hermitian systems induced
by hopping anisotropy and random disorder potential. To this end, we compute 
the disorder-averaged inverse participation ratio 
(IPR)~\cite{visscher_72,evers_00}. It is worth noting that for non-Hermitian 
systems, the IPR can be defined in two ways: (i) for the right (left) 
eigenstates of the model Hamiltonian $\hat{H}$ ($\hat{H}^{\dagger}$) and (ii) 
for the biorthogonal basis using both right and left 
eigenstates~\cite{xiao_22}. The disorder-averaged IPR of the $n$th right 
eigenstate and biorthogonal eigenstates are 
\begin{eqnarray}
  I_{n} = \frac{\sum_{i} \overline{|\phi^{(s,n)}_{i}|^{2}}}
			 {(\sum_{i}|\phi^{(s,n)}_{i}|)^{2}}; ~~~
  I_{nB} = \frac{\sum_{i} \overline{|\tilde{\phi}^{(s,n)}_{i}|^{2}}}
			 {(\sum_{i}|\tilde{\phi}^{(s,n)}_{i}|)^{2}},
\end{eqnarray}
where the overline indicates the average over disorder realizations with 
corresponding index $s$, 
$\phi^{(s,n)}_{i}\equiv(\langle n| b_{i}\rangle_{s}^{*})\langle 
n| b_{i}\rangle_{s}$ and
$\tilde{\phi}^{(s,n)}_{i}\equiv(\langle \tilde{n}| b_{i} \rangle_{s}^{*})
\langle n| b_{i} \rangle_{s}$ with $n$ and $\tilde{n}$ are the right and
corresponding left eigenstates, and $|b_{i}\rangle$ are Fock space basis. The 
mean IPR $\overline{I}=\sum_{n} I_{n}/D_{H}$ and mean biorthogonal IPR 
$\overline{I}_{B}=\sum_{n} I_{nB}/D_{H}$ are obtained by averaging over the 
whole energy spectrum. Here, $D_{H}$ is the dimension of the Hilbert space of 
many-body Hamiltonian. For localized states, IPR approaches a finite value 
(unity) while it approaches zero for delocalized states in the thermodynamic 
limit. The biorthogonal IPR includes the nonorthogonality of different 
eigenstates, thus captures the non-Hermitian effects of interacting disordered 
systems~\cite{wang_19,suthar_22}. 
\begin{figure}[ht]
  \includegraphics[width=\linewidth]{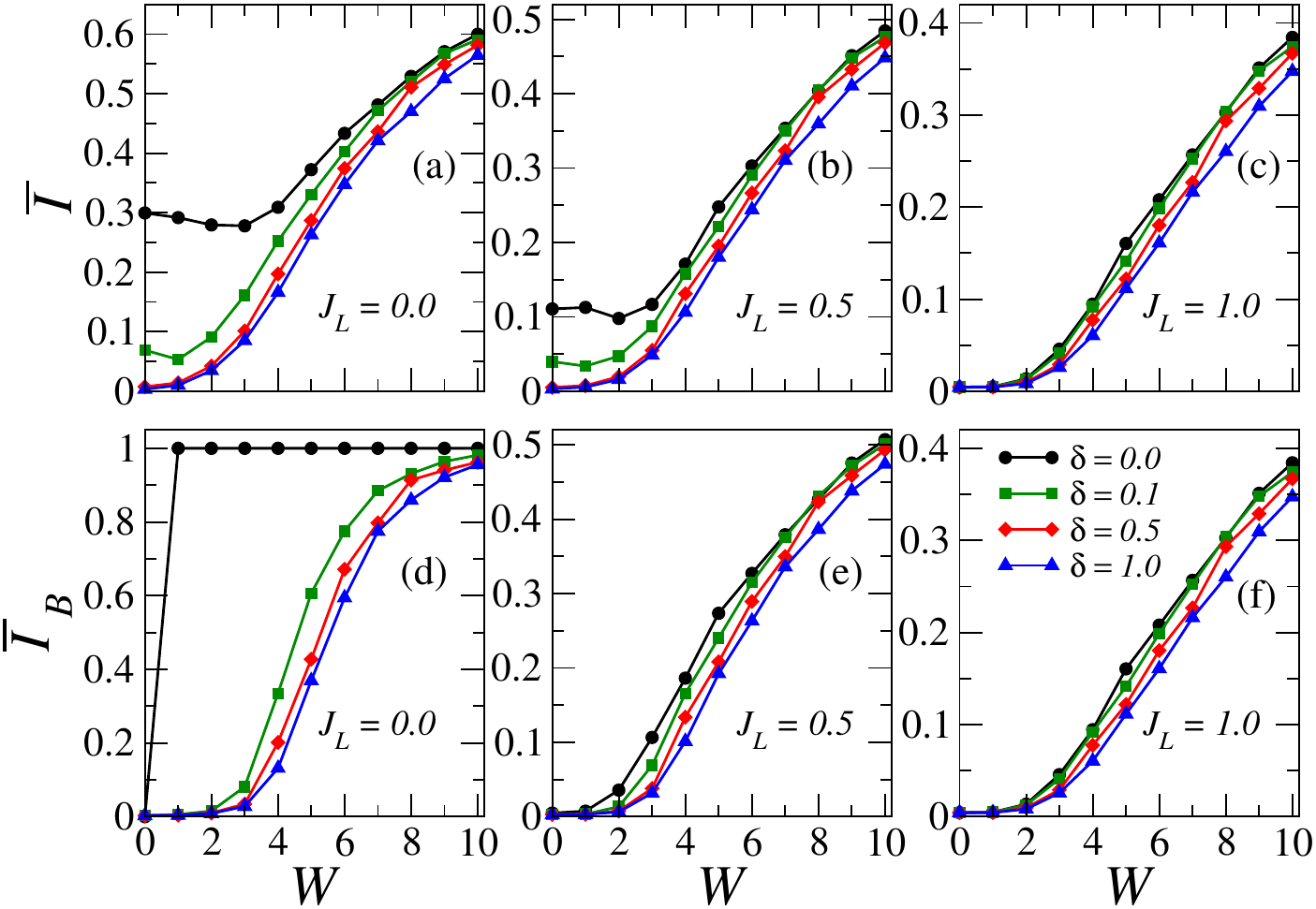}
  \caption{The disorder-averaged IPR and the biorthogonal IPR for system size 
	   $L=12$. (a,d) The non-reciprocal case with unidirectional hopping 
	   $J_{L}=0$, (b,e) The non-reciprocal case with $J_{L}=0.5$, and (c,f) 
	   $J_{L}=1$ corresponds to the Hermitian limit. The upper (a,b,c) 
	   and lower (d,e,f) panel shows IPR and biorthogonal IPR, 
	   respectively. The black, green, red, and blue colored lines 
	   represent the cases with boundary parameters $\delta=0, 0.1, 0.5,$ 
	   and $1$, respectively. The IPR values are obtained by averaging the 
	   energy spectrum. Here, $J_{R}=1$ and data is averaged over $300$ 
	   disorder realizations.}
\label{ipr}
\end{figure}
\begin{figure}[ht]
  \includegraphics[width=\linewidth]{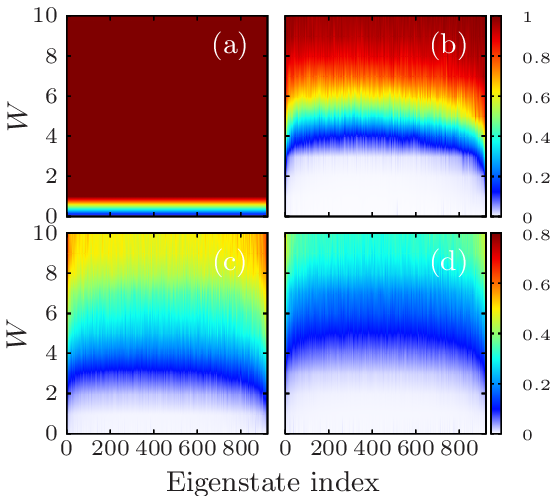}
  \caption{The disorder-averaged biorthogonal IPR in the whole energy spectrum
           as a function of $W$ for system size $L=12$. Some specific cases
           are considered: (a) for unidirectional hopping with OBC $(\delta=0)$
           case of generalized boundary, (b) for unidirectional hopping with
           $\delta=0.5$, (c) $J_{L}=0.5$ with OBC, and (d) $J_{L}=\delta=1$
           corresponding to Hermitian limit with  PBC. The spectrum of 
	   biorthogonal IPR is averaged over $100$ disorder realizations.}
\label{eig_bipr}
\end{figure}

In the unidirectional hopping $(J_{L}=0)$, Fig.~\ref{ipr}(a,d) shows the 
$\overline{I}$ and $\overline{I}_{B}$ as a function of $W$ for various boundary 
parameter values. At $\delta=0$ (OBC), the $\overline{I}$ is finite at weak 
disorder, suggesting the existence of non-Hermitian skin effect where the bulk 
states tend to localize at one end of a one-dimensional chain. $\overline{I}$ 
further increases with enhancement in the disorder strength $W$. The 
corresponding $\overline{I}_{B}$ approaches unity at an infinitesimal $W$. 
Since the degree of non-hermiticity is largest in the unidirectional hopping 
case, the $\overline{I}_{B}$ remains one, and the NHSE dominates over the 
disorder-induced localization. At the finite value $\delta=0.1$, the weak link 
of the ends lowers the $\overline{I}$ value (compared to the OBC case); 
however, it remains finite. At higher boundary coupling strength $\delta=0.5$, 
the $\overline{I}$ becomes zero at lower strength $W$ and increases with $W$, 
similar to the behaviour depicted at $\delta=1$ (PBC). For later three cases 
(except OBC), the biorthogonal IPR $\overline{I}_{B}$ is smaller (larger) at 
weak (strong) disorder strengths compared to $\overline{I}$. The biorthogonal 
IPR as a function of $W$ in the whole energy spectrum under the open boundary 
condition i.e. $\delta=0$ is shown in Fig.~\ref{eig_bipr}(a). The $I_{nB}$ of 
all eigenstates becomes one, demonstrating the localization due to the interplay
of skin effect and random disorder. Fig.~\ref{eig_bipr}(b) shows $I_{nB}$ at 
$\delta=0.5$ and $J_{L}=0$. With finite $\delta$, the delocalized regime 
appears as the skin-localization is suppressed and the gradual increase in the 
value of $I_{nB}$ with $W$ exhibits the disorder-driven localization. 

 For the non-reciprocal case with $J_{L}=0.5$, the values of IPR and 
biorthogonal IPR shown in Fig.~\ref{ipr}(b,e) indicates that the localization 
is sensitive to $\delta$ values. In particular, at smaller $W$, the 
$\overline{I}$ is higher due to skin effect induced localization. Moreover, the
steady $\overline{I}$ in this case is lower than Fig.~\ref{ipr}(a) as in the 
later case the unidirectional hopping promotes the skin-localization. At strong
$W$, the apparent distinctions between the behaviour of both IPRs disappear as 
the degree of non-hermiticity reduces. This is also evident from the 
distributions of $I_{nB}$ shown in Fig.~\ref{eig_bipr}(c) at $\delta=0$ 
[cf.Fig.~\ref{eig_bipr}(b)]. Finally, in the Hermitian limit $J_{L}=J_{R}=1$, 
both IPRs overlap as Hermitian systems do not show boundary sensitivity and are
free from NHSE. The increase in IPRs represents the initially extended states 
get localized by the growing disorder. This behaviour for the Hermitian system 
is depicted in Fig.~\ref{ipr}(c,f). The distinct localization of eigenstates 
solely due to disorder potential is presented in Fig.~\ref{eig_bipr}(d).
\begin{figure}[ht]
  \includegraphics[width=\linewidth]{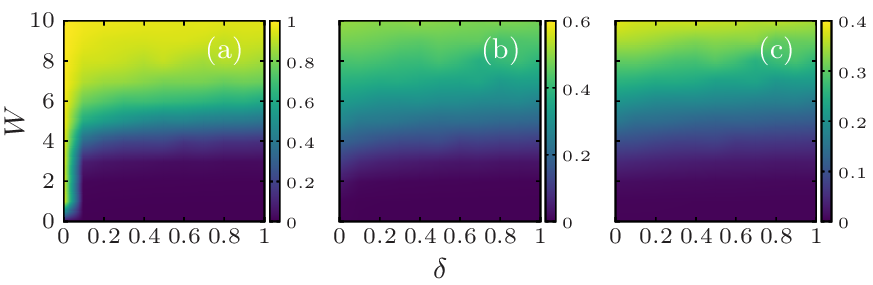}
  \caption{The biorthogonal IPR is a function of boundary parameter and 
	   disorder strength. Here $J_R=1$ and $J_L$ is varied to have effects
	   from non-reciprocal coupling to the Hermitian limit. (a) $J_L = 0$, 
	   (b) $J_L = 0.5$, and (c) $J_L = 1.0$. The values of IPR are obtained
	   by averaging over $300$ disorder realizations.}
\label{bipr_delta}
\end{figure}

Next we turn to study the localizations using biorthogonal IPR as a function of
$\delta$ and $W$. The averaged $I_{nB}$ for three cases: unidirectional, 
non-reciprocal, and reciprocal hoppings are shown in Fig.~\ref{bipr_delta}. At 
$J_{L}=0$, when $\delta=0$, the particles tend to localize at one end and thus 
leads to finite $I_{nB}$ (unity) even at smaller disorder strengths 
[Fig.~\ref{bipr_delta}(a)]. This behaviour is consistent with 
Fig.~\ref{ipr}(a). The introduction of a small link or finite $\delta$ departs 
from the OBC case, and thus $I_{nB}$ gradually increases with $W$ and becomes 
unity. On the contrary, for a nonzero left hopping as shown in 
Fig.~\ref{bipr_delta}(b), the $I_{nB}$ is zero at smaller disorder strengths 
for $\delta=0$. In Hermitian limit at $J_{L}=1$, the value of $I_{nB}$ lowers 
at strong disorder as the localization is due to the disorder potential, and 
NHSE does not contribute. 

%%%%%%%%%%%%%%%%%%%%%%%%%%%%%%%%%%%%%%%%%%%%%%%%%%%%%%%%%%%%%%%%%%%%%%%%%%%%%%%
%%%%                         C. Level statistics                           %%%%
%%%%%%%%%%%%%%%%%%%%%%%%%%%%%%%%%%%%%%%%%%%%%%%%%%%%%%%%%%%%%%%%%%%%%%%%%%%%%%%
\subsection{Level statistics}
\label{level_stat}
\begin{figure}[ht]
  \includegraphics[width=\linewidth]{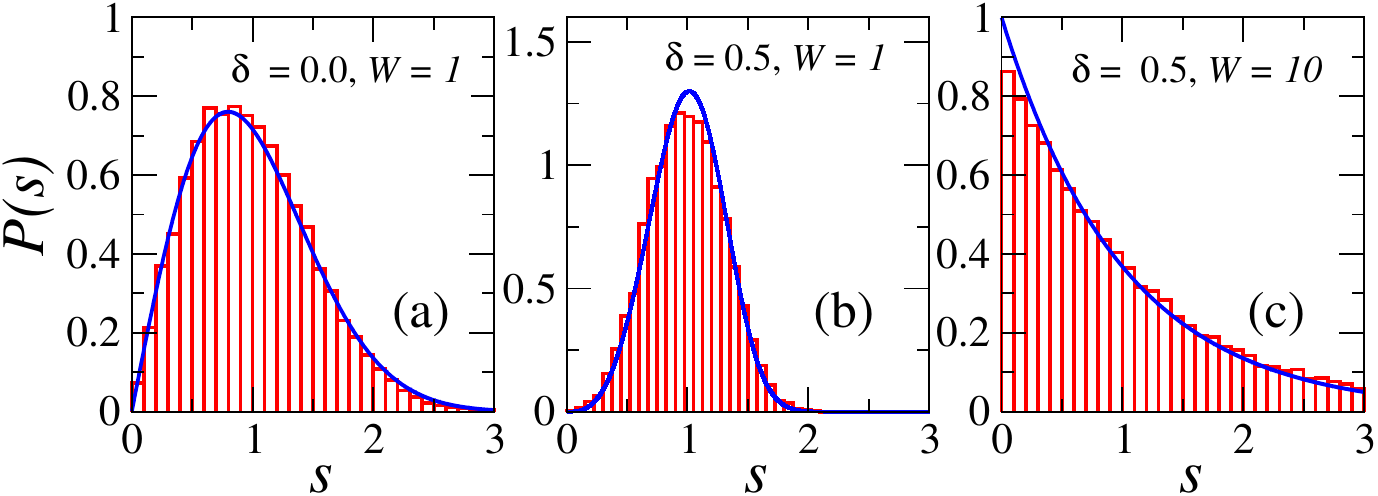}
  \caption{The level-spacing distributions for non-reciprocal hopping strengths 
	   lead to non-hermiticity of the system. (a,b) At weak disorder, 
	   the boundary coupling exhibits a transition of nearest level-spacing 
	   distribution from GOE to Ginibre distribution, corresponding to the 
	   real-complex transition. (c) At strong disorder the 
	   non-Hermitian MBL regime corresponds to the Poisson distribution. 
	   Statistics are taken for eigenstates lying within $10\%$ of the
	   middle of the spectrum in the complex energy plane. The solid blue
	   line represents the (a) GOE, (b) Ginibre, and (c) Poisson 
	   distributions. Here, the non-reciprocal hoppings are $J_{L}=0.5$ 
	   and $J_{R}=1.0$. The system size is $L=16$.}
\label{dist}
\end{figure}

We now proceed to study the level statistics of the complex energy spectrum to 
understand the localization properties corresponding to the real-complex 
transitions of the model~\cite{bohigas_84,guhr_98,mehta_04}. We consider the 
nearest-level spacing $d_{1,i} \equiv {\rm min}_{j} |E_i - E_j|$ which defines 
the minimum (nearest-neighbour) distance between two eigenenergies $E_i$ and 
$E_j$ in the complex plane. We employ the unfolding procedure of the complex 
and real energy spectrums to obtain the nearest-neighbour 
distance~\cite{haake_13,hamazaki_20,shivam_23}. The local mean density of the 
energy eigenvalues is 
\begin{equation}
  \bar{\rho}_{i} = n/(\pi d^{2}_{n,i}), 
\end{equation}
where $n$ is chosen to be very small compared to the Hamiltonian matrix size 
and sufficiently larger than one (approximately $30$). $d_{n,i}$ is the $n$th 
nearest-neighbour distance from $E_{i}$. After removing the dependence of 
local density on the level spacing, the rescaled nearest-neighbour distance is 
\begin{equation}
  s_{i} = d_{1,i}~\sqrt{\bar{\rho}_{i}}. 
\end{equation}
The probability distribution of $\{s_{i}\}$ denoted as $P(s)$ is plotted in 
Fig.~\ref{dist} for $\delta=0$ and $\delta=0.5$ at weak disorder, and 
$\delta=0.5$ at strong disorder. With non-reciprocal hopping, at $\delta=0$ and
$W=1$, the system is in the delocalized phase with real energy spectrum and the
level-spacing distribution approaches the distribution of Gaussian orthogonal 
ensemble (GOE)~\cite{grobe_88}, which is given by 
\begin{equation}
  P^{\rm R}_{\rm GOE}(s) = \frac{\pi s}{2}~\exp\left(\frac{-\pi~s^2}{4}\right).
\end{equation}
The corresponding distribution is presented in Fig.~\ref{dist}(a) for 
non-reciprocal hoppings, $J_{L}=0.5$ and $J_{R}=1$. Note that at weak disorder 
with $\delta=0$, both IPR and biorthogonal IPR differ in values, and the later 
is zero as it is not suffered from NHSE (cf. Fig.~\ref{ipr}(b)). As discussed 
earlier, an infinitesimal boundary coupling leads to a real-complex transition 
at weak disorder strengths. This transition results in to change in the 
distributions from that of GOE to the Ginibre 
ensemble~\cite{ginibre_65,grobe_88,mehta_04,haake_13}. 
As an illustration, for $\delta=0.5$ and $W=1$, the system is in delocalized 
phase with complex energies, and it exhibits Ginibre distribution 
[Fig.~\ref{dist}(b)] $P^{\rm C}_{\rm Gin}(s)=cp(cs)$, which characterizes an 
ensemble for non-Hermitian Gaussian random matrices. Here, 
\begin{equation}
  p(s) = \lim_{N\rightarrow\infty}
         \left[\prod_{n=1}^{N-1}e_n(s^2)e^{-s^2}\right]
         \sum_{n=1}^{N-1}\frac{2s^{2n+1}}{n!e_n(s^2)}
\end{equation}
with $e_n(x)=\sum_{m=0}^n\frac{x^m}{m!}$ and 
$c=\int_0^\infty ds~s~p(s)=1.1429$~\cite{grobe_88,markum_99}.
At $W=10$, the system is in MBL phase, and the eigenspectrum becomes real. 
Since the MBL phase persists even in the presence of non-reciprocal hoppings, 
so dubbed non-Hermitian MBL. In MBL phase, the real eigenenergy spectrum at 
strong disorder follows the real Poisson distribution $P^{\rm R}_{\rm Po}(s)=
\exp(-s)$. This is evident from Fig.~\ref{dist} (c). The level statistics can 
demarcate the MBL phase by showing the characteristic level distributions. 

We further consider the complex level-spacing 
ratio~\cite{lucas_20,suthar_22,xiao_22a} for $i$th eigenvalue, which is a 
dimensionless complex variable and defined as
\begin{equation}
  z_i = \frac{E_{i} - E^{\rm NN}_{i}}{E_{i} - E^{\rm NNN}_{i}} 
\equiv r_i e^{i\theta_{i}} 
\end{equation}
with the amplitude $r_i \equiv|z_i|$. The $E^{\rm NN}_{i}$ and 
$E^{\rm NNN}_{i}$ are the nearest and next-nearest neighbours of the energy 
level $E_{i}$ in the complex plane, respectively. In general, the 
nearest-neighbour distance is not universal and depends on the local density of
states. However, in the ratio $z_i$, the dependence of the local density is 
washed away. This ratio is an ideal diagnostic to examine the ergodicity to MBL
transition. The mean level-spacing ratio $\langle r\rangle$ is obtained 
by the average of $r_i$ over the energy window and number of disorder 
realizations. In the present study, we consider the energy window to be $10\%$ 
energy eigenvalues around the center of the eigenspectrum in the complex energy
plane. This choice allows us to obtain a large number of eigenvalues for the 
level statistics analysis and ascertain that their eigenstates share similar 
localization properties. The number of disorder realizations is chosen such 
that the total number of eigenvalues is $10^{6}$.
\begin{figure}[ht]
  \includegraphics[width=\linewidth]{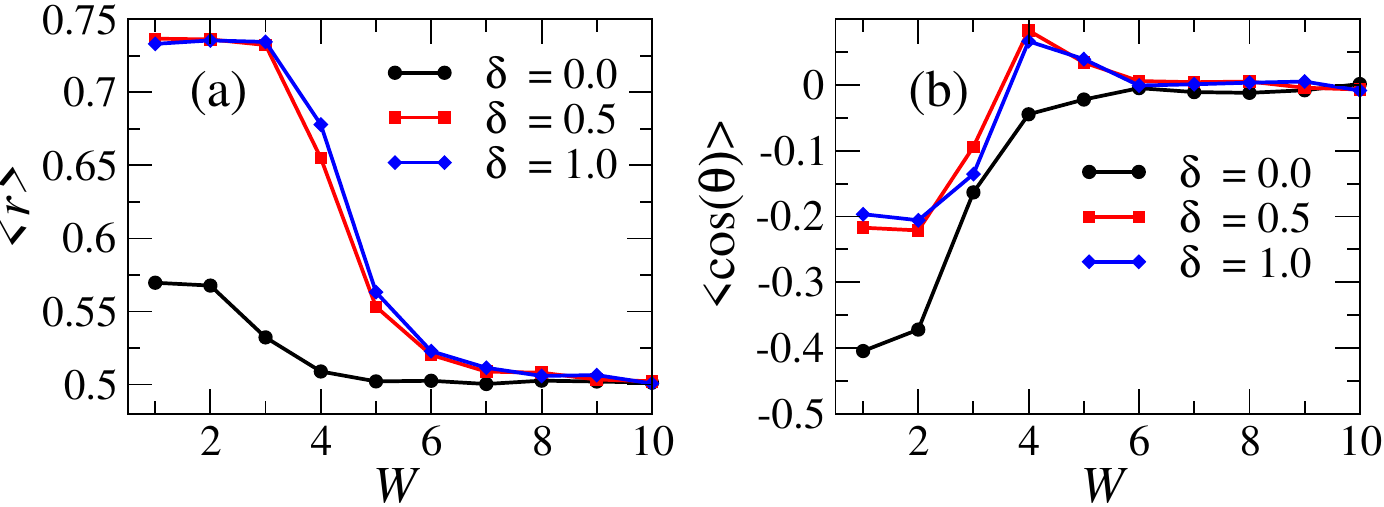}
  \caption{(a) The average complex level-spacing ratio and (b) the cosine as a 
	   function of boundary parameters $\delta$'s. An infinitesimal change 
	   in boundary parameter results in a change in the level statistics. 
	   We consider the system size $L=16$. The level-spacing ratio is 
	   obtained for eigenstates lying within $10\%$ of the middle of the 
	   eigenspectrum. The disorder average is performed for $100$ 
	   realizations.}
  \label{rbar-cos}
\end{figure}

We first consider the evolution of averaged $\langle r\rangle$ as a
function of $W$, as shown in Fig.~\ref{rbar-cos}(a) for $L=16$. It is shown for
non-reciprocal hoppings $J_{L}=0.5$ and $J_{R}=1$. For $\delta=0$, when $W$ is 
smaller, $\langle r\rangle$ attains a constant value of
$0.569$~\cite{peron_20,xiao_22a} for the GOE distribution [Fig.~\ref{dist}(a)].
As $W$ increases, the value of $\langle r\rangle$ decreases and stabilizes at 
strong disorder. The value of $\langle r\rangle$ at strong disorder is $0.5$, 
corresponding to real Poisson level statistics~\cite{peron_20,xiao_22a}. Thus, 
even with real spectrum at $\delta=0$, the level-spacing distribution and 
spacing ratio $\langle r\rangle$ can distinguish delocalization-localization 
transition driven by random disorder. Note that at smaller $W$, the reality of 
the eigenspectrum is due to OBC, as the time-reversal symmetric model 
Hamiltonian can be mapped to the corresponding Hermitian model using imaginary 
gauge transformation~\cite{hatano_21,suthar_22}. We next analyze the 
level-statistics with non-zero $\delta$ of the generalized boundary, where the 
system shows complex-real transition with $W$. An infinitesimal coupling of 
both ends leads to $\langle r\rangle\approx 0.74$~\cite{lucas_20,antonio_22} at
a smaller $W$ for the Ginibre distribution [Fig.~\ref{dist}(b)]. At strong $W$,
the $\langle r\rangle$ is $0.5$. The same transition of $\langle r\rangle$ is 
also noted for the PBC case with $\delta=1$. The change in value of 
$\langle r\rangle$ is consistent with the corresponding complex-real transition
shown in Fig.~\ref{imag-frac}(b).

Similarly, a variable related to the angular distributions of complex spacing 
ratio $\langle\cos\theta\rangle$~\cite{lucas_20,antonio_22} is also a 
single-number signature that distinguishes the different phase regimes with 
different level distributions. The change in the value of 
$\langle\cos\theta\rangle$ as a function of $W$ for three values of $\delta$ is
shown in Fig.~\ref{rbar-cos}(b). For $\delta=0$, the system exhibits GOE 
distributions with $\langle\cos\theta\rangle\approx -0.4$ at smaller $W$. As 
$W$ increases, the value of $\langle\cos\theta\rangle$ increases and eventually
becomes zero for Poisson statistics at strong disorder. At finite $\delta$, 
$\delta=0.5$ and $\delta=1$, the $\langle\cos\theta\rangle\approx-0.2$ for the 
complex eigenenergy phase with Ginibre distribution, and the value approaches 
zero at strong disorder strengths. Thus, both $\langle r\rangle$ and 
$\langle\cos\theta\rangle$ serve as a good indicator of localization 
transitions with change in the level distributions. The $\delta$ and $W$ 
dependence of real-complex transitions are consistent with the pertinent 
level-spacing distribution transitions. The critical disorder strength 
$W_c\approx4$ at which the transition to localization phase occurs, agrees well
with real-complex spectral transition and biorthogonal IPR 
[cf.~Fig.~\ref{rbar-cos}(b), Fig.~\ref{bipr_delta}(b), and 
Fig.~\ref{imag-frac}(b)].

%%%%%%%%%%%%%%%%%%%%%%%%%%%%%%%%%%%%%%%%%%%%%%%%%%%%%%%%%%%%%%%%%%%%%%%%%%%%%%%
%%%%%                       D. Dynamical properties                       %%%%%
%%%%%%%%%%%%%%%%%%%%%%%%%%%%%%%%%%%%%%%%%%%%%%%%%%%%%%%%%%%%%%%%%%%%%%%%%%%%%%%
\subsection{Dynamical properties}
\label{dyna_prop}
We finally examine the dynamical properties of the many-body non-Hermitian  
system using the quantum trajectory approach~\cite{daley_14,ashida_20}. For a 
given initial state $\left|\psi_{0}\right\rangle$ at $t=0$, the time evolved 
wave-function of non-equilibrium dynamics is 
\begin{equation}
  \ket{\psi_{t}} = \frac{e^{-{i}\mathcal{H}t/\hbar}\ket{\psi_{0}}}
                   {\sqrt{\bra{\psi_{0}}
                   e^{{i}\mathcal{H}^{\dagger}t/\hbar}
                   e^{-{i}\mathcal{H}t/\hbar}\ket{\psi_{0}}}},
\label{wavefunction}
\end{equation}
where $\mathcal{H}$ is an effective non-Hermitian Hamiltonian governing the 
dynamics of non-Hermitian systems. To explore the non-equilibrium dynamics, we 
choose the density-wave ordered state $\ket{101010\cdots}$ as an initial state 
where odd sites are occupied and even sites are empty. This state has routinely 
been prepared in experiments to investigate the localization properties of 
Hermitian systems~\cite{schreiber_15,kohert_19,scherg_21}.
\begin{figure}[ht]
  \includegraphics[width=\linewidth]{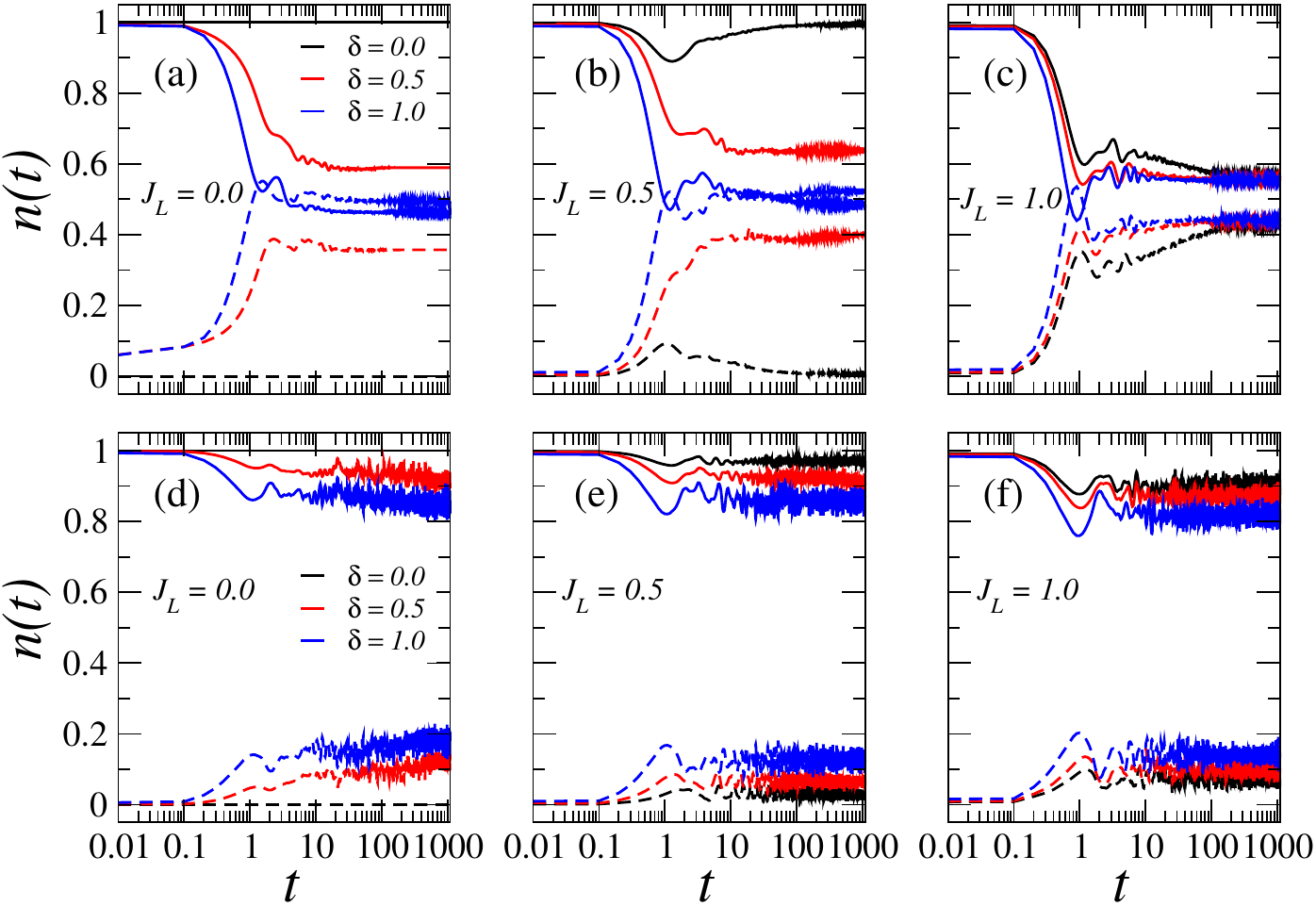}
  \caption{The time evolution of disordered-averaged local particle density 
	   for various sets of left hopping strength and boundary parameters. 
	   This is shown for two regimes: (a,b,c) weak disorder $W=2$ and 
	   (d,e,f) strong disorder $W=14$. The solid lines indicate the 
	   evolution of density at the first lattice site of the chain $n_{1}$ 
	   and the corresponding dashed lines represent the density at the 
	   right end of the chain $n_{L}$. Here, the system size is $L=16$ and 
	   $J_{R}=1.0$. The density is averaged over $100$ disorder 
	   realizations. The initial state is 
	   $\ket{\psi_{0}} = \ket{1010\cdots}$.}
  \label{den}
\end{figure}

We first examine the time dynamics of local particle density at two ends of the 
chain, i.e., the density at the first and last sites. The dynamics of local 
particle (normalized) density $n_{1}$ and $n_{L}$ are presented in 
Fig.~\ref{den} for several cases with a set of $J_{L}$ and $\delta$. To 
understand the interplay of boundary sensitivity and disorder potential, it is 
shown for weak (a,b,c) and strong disorder (d,e,f) regimes. We begin with the 
evolution of $n_{1}$ and $n_{L}$ at weak disorder $W=2$. When both $J_{L}$ and 
$\delta$ are zero, which refers to the case of an open interacting chain with 
unidirectional hopping, all the atoms accumulate at one end, leading to 
$n_{1}=1$ while $n_{L}=0$ [Fig.~\ref{den}(a)]. A small boundary coupling while 
maintaining the unidirectional hoppings results in delocalization of atoms 
occupying across the lattice sites. Moreover, a PBC case with $\delta=1$ also 
exhibits similar behaviour [Fig.~\ref{den}(a)]. A small left hopping 
$J_{L}=0.5$ with no boundary coupling leads to accumulation of density at the 
first site over a long time [Fig.~\ref{den}(b)]. Recall that the accumulation 
of atoms at the ends is due to the prevailing NHSE for OBC ($\delta=0$) cases. 
The boundary coupling leads to delocalization of atoms and atoms do not 
accumulate at the ends of the chain despite the system being non-Hermitian due 
to non-reciprocal hoppings. Finally, for $J_{L}=1$, the system is Hermitian, 
thereby even $\delta=0$ does not accumulate the particles at the ends. 
And, no boundary sensitivity is seen for $J_{L}=1$ [Fig.~\ref{den}(c)]. As the 
disorder strength is small, the delocalization of atoms is favoured, as 
apparent from the local particle densities. 

We now turn to discuss the role of disorder potential. The local particle 
density at $W=14$ is shown in Fig.~\ref{den}(d,e,f). For a unidirectional open 
chain, the role of NHSE dominates, leading to full localization of atoms at 
the first site. However, in other cases the localization is mainly due to 
strong disorder, and the effects of the boundary parameter fade away. This is 
evident from the particle densities of non-Hermitian cases 
[Fig.~\ref{den}(d,e)], which are similar to $J_{L}=1$ (Hermitian) cases
[Fig.~\ref{den}(f)]. 

We further calculate the population imbalance, a measure of disorder-induced
localization~\cite{alet_18,abanin_19}, defined as
\begin{equation}
  I(t) = \sum_{j} (-1)^{j} n_{j}.
\end{equation}
\begin{figure}[ht]
  \includegraphics[width=\linewidth]{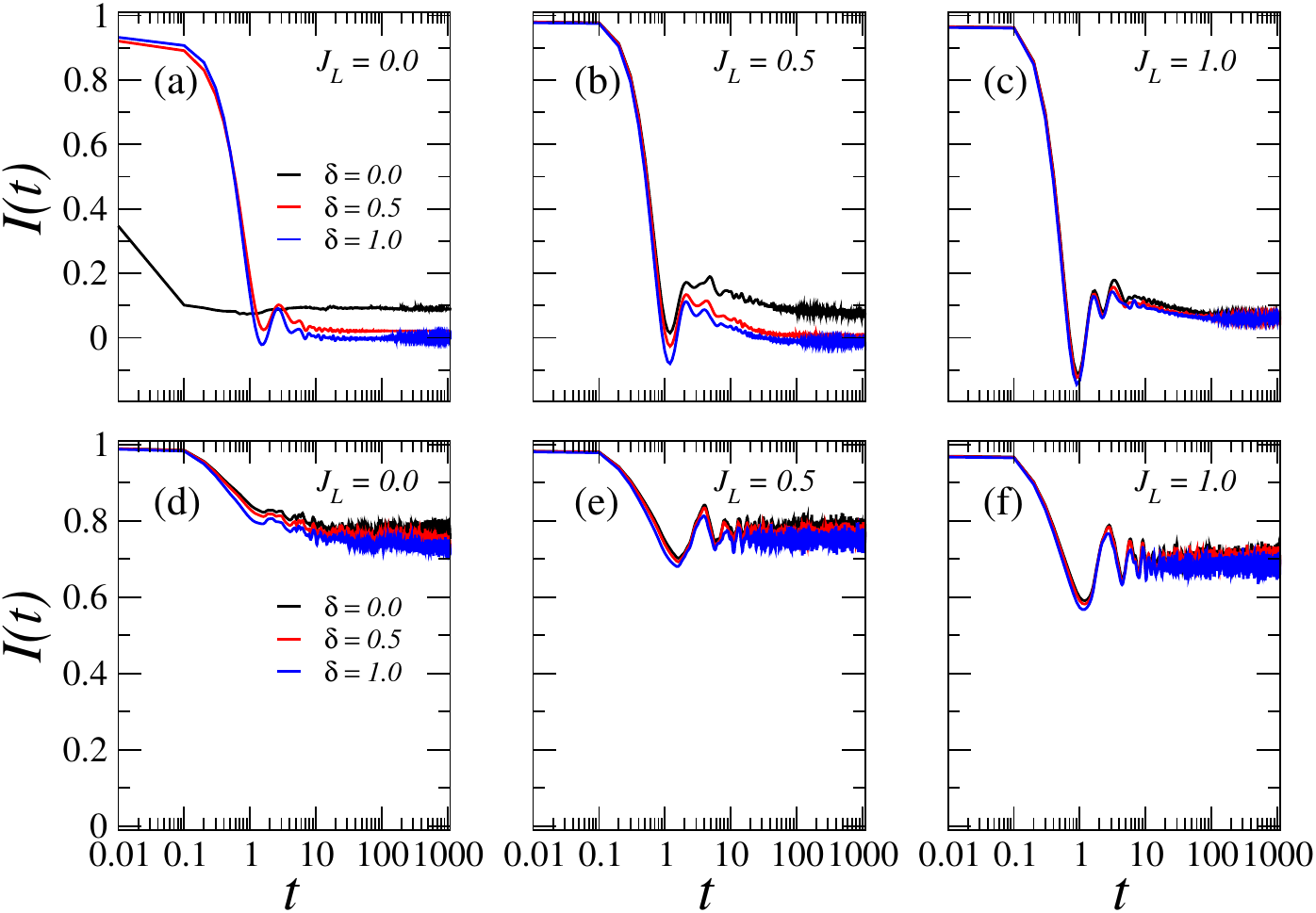}
  \caption{The time evolution of population imbalance for various sets of left 
   	   hopping strength and boundary parameters. This averaged imbalance is 
	   shown for two values of disorder strengths: (a,b,c) $W=2$ and 
	   (d,e,f) $W=14$. In the delocalized regime with weak disorder, $I(t)$ 
	   relaxes to zero while in the localized regime with strong disorder 
	   it saturates to a finite value characterizing (non-) Hermitian MBL. 
	   Here, the imbalance is averaged over $100$ samples for system size 
	   $L=16$. The right-hopping $J_{R}$ is $1.0$. The initial state is 
	   taken as $\ket{\psi_{0}} = \ket{1010\cdots}$.}
  \label{imbal}
\end{figure}
The initial density-wave ordered state maximizes the imbalance at $t=0$ for 
spinless fermions with half-filling. It is worth noting that the imbalance does
not characterize the NHSE driven localization; however, it can be identified 
based on the local particle distributions, as discussed previously. In the 
followings, we investigate the experimentally accessible time evolution of 
disordered-averaged imbalance. Fig.~\ref{imbal} displays the time evolution of 
disordered averaged imbalance for the initially prepared density-wave ordered 
state. The dynamics are shown for $W=2$ (a,b,c) and $W=14$ (d,e,f), 
corresponding to extended and MBL regimes, respectively. Let us recall that the
memory of the initial state is lost at small disorder strength that signals the 
delocalization of the system and decay of imbalance. While the MBL phase is 
characterized by a finite stationary value acquired by imbalance at long-time 
evolution. In the extended regime $(W=2)$, the imbalance decays to zero for all
non-Hermitian and Hermitian cases considered, which implies the relaxation of 
the initial density profile regardless of the values of $J_{L}$ and $\delta$. 
However, the NHSE-driven localization in weak disorder regime is concluded for 
non-Hermitian OBC cases [Fig.~\ref{den}(a,b)]. 

For strong disorder, $W=14$, on the other hand, the imbalance saturates to 
a finite value, signifying the property of MBL phase. Hence, for all cases, the
strong disorder leads to localization. This is consistent with the 
eigenspectrum [Fig.~\ref{imag-frac}], where the MBL phase at strong disorder 
suppresses the imaginary part of the complex energies (dynamically stabilizes) 
and results in a real eigenspectrum. 

\begin{figure}[ht]
  \includegraphics[width=\linewidth]{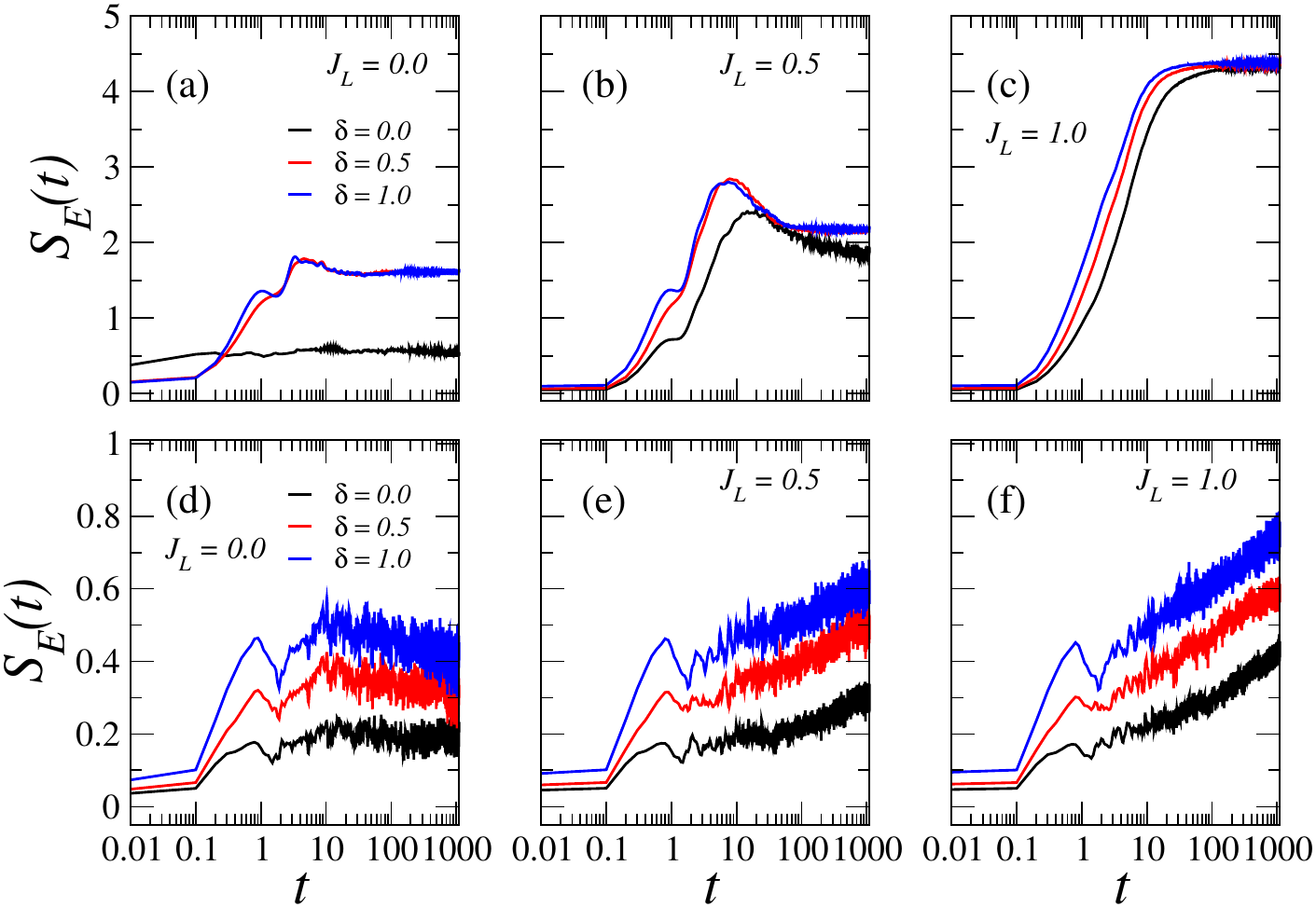}
  \caption{The dynamics of half-chain entanglement entropy for different values
	   of $\delta$ with (a,d) $J_{L}=0$, (b,e) $J_{L}=0.5$, and 
	   (c,f) $J_{L}=1.0$ at weak disorder (upper panel) and strong disorder
	   (lower panel). (a,b,c) At $W=2$, $S_{E}(t)$ of an initial 
	   density-wave ordered state $\ket{1010\cdots}$ first increases 
	   linearly and then saturates for reciprocal hopping cases while it 
	   first increases and then decreases before attaining a steady value 
	   for non-reciprocal hoppings. (d,e,f) At $W=14$, $S_{E}(t)$ exhibits 
	   logarithmic growth for reciprocal and non-reciprocal hoppings. The 
	   results are averaged over $100$ disorder realizations for system 
	   size $L=16$.}
  \label{entang}
\end{figure}

We finally investigate the growth of the entanglement entropy to corroborate 
the role of hopping non-reciprocity of spinless fermions with generalized 
boundary. The half-chain entanglement entropy~\cite{alet_18,abanin_19} is 
defined as 
\begin{equation}
  S_{E}(t) = - \Tr\left[\rho_{A}(t) \ln \rho_{A}(t)\right],
\end{equation}
where the two subsystems are denoted as $A$ and $B$ with 
$\rho_{A}(t) = \Tr_{B}{\ket{\psi_{t}}\bra{\psi_{t}}}$ being the reduced density
matrix of subsystem $A$. Here, $\Tr_B$ is the trace over degrees of freedom of 
subsystem $B$. The entanglement entropy of the initial wave-packet with 
non-reciprocal hopping does not spread as in the Hermitian case (in particular
at weak disorder), rather slides as determined by the asymmetry in the 
hopping~\cite{hamazaki_19,suthar_22,orito_22,wang_23,orito_23}.  

The time evolutions of disorder-averaged $S_{E}(t)$ at weak and strong disorder 
strengths are shown in Fig.~\ref{entang}. For unidirectional hopping under open 
boundaries, localization leads to lower entanglement entropy. With boundary 
terms, the coupling of end sites results in higher entanglement entropy and 
over long time it remains steady. For weak disorder, with non-reciprocal 
hopping the entanglement entropy decreases after a transient increase and 
attains a steady value over time [Fig.~\ref{entang}(a,b)]. For the Hermitian 
system, the entanglement entropy grows rapidly with time and saturates to 
larger values, a feature expected for the extended phase 
[Fig.~\ref{entang}(c)]. 

In the strong disorder regime, the entanglement entropy can be described as a 
logarithmic function of time, compatible with the characteristic of MBL phase
[Fig.~\ref{entang}(e,f)]. Hence, the entropy of entanglement of initially 
separable density-wave ordered states grows logarithmically in time after 
initial transients signify MBL with non-reciprocal hopping under generalized 
boundary.

%%%%%%%%%%%%%%%%%%%%%%%%%%%%%%%%%%%%%%%%%%%%%%%%%%%%%%%%%%%%%%%%%%%%%%%%%%%%%%%
%%%%                       IV. Conclusions                                 %%%%
%%%%%%%%%%%%%%%%%%%%%%%%%%%%%%%%%%%%%%%%%%%%%%%%%%%%%%%%%%%%%%%%%%%%%%%%%%%%%%%
\section{Conclusions}
\label{conc}
 We solved exactly the non-Hermitian Hatano-Nelson model of interacting 
spinless fermions and investigated the interplay of localization phenomena due 
to boundary sensitivity and disorder potential. An infinitesimal variation in 
the boundary term leads to spectrum and localization transitions in a 
one-dimensional fermionic chain. The random disorder potential washes away the 
real-complex transitions due to the prevailing non-Hermitian MBL. We further 
analyze IPR with its biorthogonal version and level statistics to reveal the 
coincidence of real-complex spectral transition with localization transition. 
Moreover, the emergence of real spectra due to infinitesimal change in boundary
parameter even at weaker disorder is attributed to NHSE, which is identified 
using nearest-neighbour level spacing distributions. Finally, the two 
localization phenomena, namely NHSE and MBL, are distinguished by the 
non-equilibrium dynamics of local particle densities and population imbalance. 
The time evolution of densities demonstrates the atomic localizations at the 
first site while that of imbalance shows relaxation to a finite value. This 
confirms NHSE-driven localization with non-reciprocal hopping. While the steady
imbalance and logarithmic growth of entanglement entropy at strong disorder 
validate the features of MBL in interacting non-Hermitian systems. With the 
recent advancement in engineering non-reciprocal hopping by reservoir coupling, 
we believe that our findings will inspire the realization of localization 
transitions in non-Hermitian disordered systems.

%%%%%%%%%%%%%%%%%%%%%%%%%%%%%%%%%%%%%%%%%%%%%%%%%%%%%%%%%%%%%%%%%%%%%%%%%%%%%%%
%%%%                           Acknowledgments                           %%%%%%
%%%%%%%%%%%%%%%%%%%%%%%%%%%%%%%%%%%%%%%%%%%%%%%%%%%%%%%%%%%%%%%%%%%%%%%%%%%%%%%
\begin{acknowledgments}
  We thank Yi-Cheng Wang, Jhih-Shih You, and H. H. Jen  for valuable 
discussions. K.S. acknowledges support from the Science and Engineering 
Research Board, Department of Science and Technology, Government of India 
through Project No. SRG/2023/001569.
\end{acknowledgments}

\bibliography{bndry_dord}
\end{document}